\documentclass[12pt]{article}
\usepackage{amsmath,epsfig,euscript}

\def\e{\varepsilon}

\begin{document}

\begin{titlepage}

\begin{flushright}
MIT-CTP~3695\\
{\tt hep-ph/0511098}\\[0.2cm]
November 8, 2005
\end{flushright}

\vspace{0.7cm}
\begin{center}
\Large\bf\boldmath
Shape-function independent relations of charmless inclusive $B$-decay spectra
\unboldmath
\end{center}

\vspace{0.8cm}
\begin{center}
{\sc Bj\"orn O. Lange}\\
\vspace{0.4cm}
{\sl Center for Theoretical Physics\\ 
Massachusetts Institute of Technology\\
Cambridge, MA 02139, U.S.A.}
\end{center}

\vspace{1.0cm}
\begin{abstract}
\vspace{0.2cm}
\noindent 
A leading-power factorization formula for weight functions relating
the $\bar B\to X_s\gamma$ photon spectrum to arbitrary partial decay
rates in $\bar B\to X_u\,l^-\bar\nu$ is derived. These weight
functions are independent of the hadronic shape function and allow for
the determination of $|V_{ub}|$ in a model-independent way. We
calculate the weight function in renormalization-group improved
perturbation theory to complete next-to-next-to leading order at the
jet scale $\mu_i^2 \sim m_b \Lambda_{\rm QCD}$ and to next-to leading
order at the hard scale $\mu_h \sim m_b$. First-order power
corrections are also included, where a model-dependence enters via the
appearance of subleading hadronic shape functions.
\end{abstract}
\vfil

\end{titlepage}

\tableofcontents

\section{Introduction}

The determination of the Cabibbo-Kobayashi-Maskawa (CKM) matrix
element $|V_{ub}|$ from inclusive semileptonic $\bar B\to
X_u\,l^-\bar\nu$ decays requires theoretical predictions for partial
decay rates, which are then compared to their experimentally measured
values. Because of a dominating $b\to c$ background in a large portion
of phase space, this procedure is adopted for a variety of restricted
regions in phase space as obtained by, e.g., accepting only events
with one or more of the following features: charged-lepton energy $E_l
\ge E_0$, hadronic invariant mass $M_X \le M_0$, leptonic invariant
mass $q^2 \ge q_0^2$, hadronic $P_+ \le \Delta$. Here, $P_\pm = E_X
\mp |\vec P_X|$, where $E_X$ denotes the energy and $\vec P_X$ the
three-momentum of the final hadronic state in the $B$-meson rest
frame. For many of these cuts a hierarchy of energy scales exists in
the decay process, for example $P_+ \sim \Lambda_{\rm QCD} \ll M_X
\sim \sqrt{m_b \Lambda_{\rm QCD}} \ll m_b$, and shape-function effects
become important \cite{Neubert:1993ch,Neubert:1993um,Bigi:1993ex}. The
theoretical expressions for differential decay rates in this region
of phase space factorize into hard functions at the scale $\mu_h\sim
m_b$, and the convolution of jet functions and shape functions at the
scale $\mu_i \sim \sqrt{m_b \Lambda_{\rm QCD}}$
\cite{Bauer:2003pi,Bosch:2004th}. While the jet functions are
perturbatively calculable, shape functions are non-perturbative
objects that capture all strong-interaction effects below the scale
$\mu_i$. At leading power the jet function $J$ and shape function
$\hat S$ are universal, and enter the QCD-factorization theorems for
both the triple differential decay rate in $\bar B\to X_u\,l^-\bar\nu$
decays and in the $\bar B\to X_s\gamma$ normalized photon spectrum
\cite{Neubert:2004dd,Neubert:2005nt}.

One strategy for the inclusive determination of $|V_{ub}|$ is to use
the $\bar B\to X_s\gamma$ photon spectrum to extract the leading shape
function $\hat S$, which then allows for the calculation of arbitrary
semileptonic partial decay rates \cite{Lange:2005yw}. It was shown in
that reference that the factorization approach can be applied to all
commonly used kinematic cuts. In practice this program is realized by
adopting a parameterizable model for the shape function, and fitting
the parameters using information of the measured photon spectrum. It
was emphasized in \cite{Lange:2005yw} that such a model is only
acceptable if the values for parameters are stable when fitted to
different aspects of the photon spectrum, such as moments of it, or
its functional form (provided that the resolution is coarse enough to
smear out hadronic resonances). With improving data on the photon
spectrum such an approach might require further and further refinement
of the models used.

A different strategy is to eliminate the necessity for the extraction
of the leading shape function, and to use the experimental data
directly. Such ideas have been investigated previously in
\cite{Neubert:1993um,Leibovich:1999xf,Leibovich:2000ey,Hoang:2005pj,%
Lange:2005qn} for some specific cuts, where partial rates in
semileptonic decays are expressed as weighted integrals over the $\bar
B\to X_s\gamma$ photon spectrum,
\begin{equation} \label{eq:SFfree}
\Gamma_u \Big|_{\rm cut} = |V_{ub}|^2 \int_0^{\Delta} dP_+\;
  W(\Delta_,P_+) \; \frac{1}{\Gamma_s(E_*)} \frac{d\Gamma_s}{dP_+}\;
\quad +\quad |V_{ub}|^2\;\Gamma_{\rm rhc} \Big|_{\rm cut}\;.
\end{equation}
In the most recent analysis of this type, \cite{Lange:2005qn}, the cut
was chosen as $P_+ \le \Delta$ for simplicity, and the weight function
$W(\Delta, P_+)$ was calculated at complete two-loop order at the
intermediate scale $\mu_i$. The second term on the right-hand side of
this equation denotes a residual hadronic power correction (rhc),
which was absorbed into the weight function in that reference. In the
above relation $\Gamma_u$ denotes the partial semileptonic decay rate
and $(1/\Gamma_s(E_*))\,(d\Gamma_s/dP_+)$ the normalized photon
spectrum in radiative decays, where $P_+ = M_B-2E_\gamma$, $E_\gamma$
is the photon energy in the $B$-meson rest frame, and the total decay
rate $\Gamma_s(E_*)$ is defined to include all events with $E_\gamma
\ge E_* = m_b/20$.  It is beneficial to use the normalized photon
spectrum instead of the absolute spectrum, because the weight function
as defined in (\ref{eq:SFfree}) is independent of $|V_{tb}V^*_{ts}|$,
possesses a well-behaved perturbative expansion \cite{Lange:2005qn},
and because the normalized photon spectrum can be determined with
better accuracy than the absolute one \cite{Neubert:2004dd}. In order
to determine $|V_{ub}|$ from relation~(\ref{eq:SFfree}) both
$\Gamma_u$ and the normalized photon spectrum enter as experimental
input, while the weight function $W(\Delta,P_+)$ and the residual
hadronic correction $\Gamma_{\rm rhc}$ are theoretical quantities.

\begin{figure}
\begin{center}
\epsfig{file=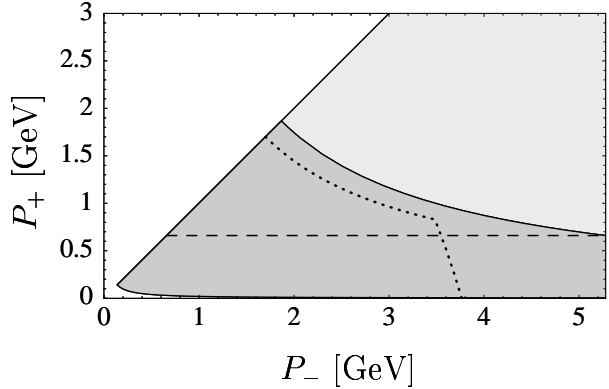, width=8cm}
\caption{\label{fig:pspace}Relevant part of the phase space of
semileptonic $B$ decays. The light-gray region denotes the portion
where a background from $\bar B \to X_c\,l^-\bar\nu$ decays exists,
while the dark-gray area is free of this background. Events with
final-state hadronic invariant mass $M_D$ are located on the line
between these two regimes. Events with $P_+ \le M_D^2/M_B$ are located
below the dashed line. As a slightly more complicated example we also
plot the borderline in phase space for events with $M_X \le 1.7$ GeV
and $q^2 \ge 8$ GeV$^2$ as a dotted line.}
\end{center}
\end{figure}

In this paper, we extend the technology developed in
\cite{Lange:2005qn} to derive the weight function for an {\em
arbitrary} kinematic cut that includes events with $P_+ \sim
\Lambda_{\rm QCD}$ and $P_- \gg \Lambda_{\rm QCD}$. In general, the
weight function $W(\Delta,P_+)$ (and the correction $\Gamma_{\rm rhc}$), 
as well as the integration limit $\Delta$ in (\ref{eq:SFfree})
depend on the particular cut that is used. For example, if we consider
a cut on the charged-lepton energy $E_l \ge E_0$ then $\Delta =
M_B-2E_0$, or if a cut on hadronic invariant mass $M_X \le M_0$ is
considered then one needs $\Delta = M_0$. A quick way to obtain some
intuition on how $W(\Delta,P_+)$ and $\Delta$ depend on the specific
cut qualitatively is to consider the $\bar B\to X_u\,l^-\bar\nu$
phase-space depicted in Figure~\ref{fig:pspace}. Note that $P_- = M_B$
in radiative decays, so that the phase-space of $\bar B\to X_s \gamma$
consists only of the vertical boundary on the right-hand side of the
plot. This picture explains correctly how the maximal $P_+^{\rm max}
\equiv \Delta$ depends on the nature of the cut, and where to expect
some kinks in $W(\Delta,P_+)$, for example for a combined cut on $M_X$
and $q^2$.

In the next section we will derive an expression for the weight
function at leading power, which follows from exact factorization
theorems for differential rates in inclusive $B$ decays in the
shape-function region. The formula for the weight function is thereby
also valid to all orders in perturbation theory. We compute its
explicit form to next-to-next-to leading order (NNLO) at the
intermediate scale $\mu_i$ and to next-to leading order (NLO) at the
hard scale $\mu_h$ in renormalization-group (RG) improved perturbation
theory, including three-loop running effects. This approximation is
sufficiently precise for phenomenological applications since effects
at subleading power become as important as higher-order perturbative
effects at leading power. In Section~\ref{sec:PowCorr} we discuss
first-order power corrections. Kinematical power corrections enter the
weight function, while hadronic power corrections give rise the
quantity $\Gamma_{\rm rhc}$ in (\ref{eq:SFfree}). We then apply our
results to a few examples of kinematic cuts in
Section~\ref{sec:examples} and perform an analysis of theoretical
uncertainties on $|V_{ub}|$ by using a simple model for the
experimental inputs.

\section{The weight function and factorization}

In this section we will adopt the leading-power approximation. The
second term on the right-hand side of relation~(\ref{eq:SFfree}) is
then absent, because it collects contributions from subleading shape
functions to both radiative and semileptonic decay rates, which start
at order $1/m_b$.

\subsection{Derivation}

We start the discussion by stating the exact factorization theorems
for the fully differential leading-power decay rate in $\bar B\to
X_u\,l^-\bar\nu$ decays
\begin{equation} \label{eq:Xulnu}
\begin{aligned}
  \frac{d^3 \Gamma_u^{(0)}}{dP_+\,dy \,d\e} 
= &\; \frac{G_F^2 |V_{ub}|^2}{192\pi^3} U(\mu_h,\mu_i) 
  (M_B-P_+)^5 y^{-2a_\Gamma(\mu_h,\mu_i)} H_u(y,\e,\mu_h) \\
& \mbox{} \times \int_0^{P_+}d\hat\omega\; 
  ym_b J(ym_b(P_+-\hat\omega),\mu_i) \,\hat S(\hat\omega,\mu_i)\;,
\end{aligned}
\end{equation}
and for the normalized $\bar B\to X_s\gamma$ photon spectrum
\begin{equation} \label{eq:Xsgamma}
\frac{1}{\Gamma_s(E_*)} \frac{d\Gamma_s^{(0)}}{dP_+} =
  \frac{U(\mu_h,\mu_i)}{H_\Gamma(E_*,\mu_h)} \frac{(M_B-P_+)^3}{m_b^3}
  \int_0^{P_+}d\hat\omega\; m_b J(m_b(P_+-\hat\omega),\mu_i) \,\hat
  S(\hat\omega,\mu_i)\;.
\end{equation}
The superscript $(0)$ indicates that these equations are valid to
leading power. In (\ref{eq:Xulnu}) $H_u(y,\e,\mu_h)$ denotes the hard
function, which collects all matching corrections at the scale $\mu_h$
and depends on the kinematic variables
\begin{equation}
y = \frac{P_--P_+}{M_B-P_+}\;,\qquad 
\e = 1- \frac{2E_l}{M_B-P_+}\;,
\end{equation}
for which the phase-space is $0\le \e \le y \le 1$. The jet function
$J(p^2,\mu_i)$ contains distributions that act on the shape function
$\hat S(\hat\omega,\mu_i)$ and depends on $y$ via $p^2 =
ym_b(P_+-\hat\omega)$. In formula~(\ref{eq:Xsgamma}) for the photon
spectrum, on the other hand, the hard function is denoted by
$H_\Gamma(E_*,\mu_h)$, and the argument of the jet function is $p^2 =
m_b(P_+-\hat\omega)$, independent of $y$ because $P_- = M_B$ in
radiative decays. Three powers of the $b$-quark mass appear due to
phase-space integrations in the total rate
$\Gamma_s(E_*)$. Renormalization-group (RG) running effects between
the scales $\mu_h$ and $\mu_i$ build up the functions $U(\mu_h,\mu_i)$
and $a_\Gamma(\mu_h,\mu_i)$ in the factorization formulas, which are
such that $U=1$ and $a_\Gamma = 0$ in the limit $\mu_i \to
\mu_h$. Expressions for them will be given below.

We start the derivation of the weight function by considering the
factorized expression for an arbitrary partial differential decay rate
\begin{equation} \label{eq:partialGamma}
  \Gamma_u^{(0)}\Big|_{\rm cut} 
= \int\limits_0^\Delta dP_+ \int\limits_0^{y_{\rm max}[P_+]}dy
  \int\limits_0^{\e_{\rm max}[P_+,y]} d\e \;
  \frac{d^3 \Gamma_u^{(0)}}{dP_+ \,dy \,d\e} \;.
\end{equation}
The integration limits $y_{\rm max}[P_+]$ and $\e_{\rm max}[P_+,y]$,
as well as $\Delta$ depend on the specifics of the cut.  (The use of
squared brackets for these quantities is supposed to remind the reader
that these functions are defined differently for different kinematic
cuts.) We are now going to rewrite this semileptonic decay rate such
that it resembles a weighted integral over the normalized photon
spectrum. Clearly the biggest obstacle is that the argument of the jet
function $J$ in (\ref{eq:Xulnu}) depends on the kinematic variable
$y$, while it does not in (\ref{eq:Xsgamma}). The solution to this
problem has been presented in \cite{Lange:2005qn}, where it was shown%
\footnote{\label{ftn:f(k)}The quantity considered in that reference was 
$f(k) = \int_0^1dy\int_0^yd\e\,y^{-2a_\Gamma} H(y,\e) Y(k,\ln y)$.}  
that a factorization of the integrated jet function in the sense that
\begin{equation} \label{eq:Ydef}
j\left(\ln \frac{m_b\Omega}{\mu_i^2} + \ln y,\mu_i\right)
= \int_0^\Omega dk\; Y(k,\ln y,\mu_i)\;
  j\left(\ln \frac{m_b(\Omega-k)}{\mu_i^2},\mu_i\right)
\end{equation}
can be achieved for arbitrary $\Omega$, if $Y(k,\ln y,\mu_i)$ is
allowed to be a distribution in the variable $k$. Here,
\begin{equation}\label{eq:j}
j\left(\ln \frac{Q^2}{\mu_i^2},\mu_i\right)
\equiv \int_0^{Q^2} dp^2\,J(p^2,\mu_i) \;.
\end{equation}
We will discuss the nature of $Y(k,\ln y,\mu_i)$ and its perturbative
expansion to two-loop order in the next section. The strategy for the
derivation of the weight function is to interchange integrations in
(\ref{eq:partialGamma}) so that the $P_+$ integration acts only on the
jet function $J$, and we can make use of (\ref{eq:Ydef}). Before this
can be done, however, it is necessary to transform the other
$P_+$-dependent terms in the partial decay rate into $P_+$-independent
ones. This can be achieved by inserting
\begin{equation}
1 = \int_0^\Delta dP'\; \delta(P'-P_+) 
  = \int_0^\Delta dP' \int_0^{M_B-P_+}dq\; \delta'(q+P'-M_B)\;,
\end{equation}
into (\ref{eq:partialGamma}), and by replacing $P_+ \to P'$ in 
$y_{\rm max}[P_+]$, $\e_{\rm max}[P_+]$ and in the kinematic prefactor
$(M_B-P_+)^5$. It is now an easy exercise to interchange the
integrations
\begin{equation}\label{eq:intchanges}
\int\limits_0^\Delta dP_+
\int\limits_0^{P_+} d\hat\omega 
\int\limits_0^{M_B-P_+} dq \ldots = 
\int\limits_0^\Delta d\hat\omega
\int\limits_0^{M_B-\hat\omega} dq
\int\limits_{\hat\omega}^{{\rm min}(\Delta,M_B-q)} dP_+ \ldots\;.
\end{equation}
Next, we apply (\ref{eq:Ydef}), interchange the $k$ and $P_+$
integrations, and undo the steps in (\ref{eq:intchanges}). Finally the
integrations over $q$ and $P'$ can be carried out, which identifies
$P' = P_++k$. As a result we arrive at an expression for the
semileptonic decay rate, which is given as an integral over the
product of the normalized photon spectrum in (\ref{eq:Xsgamma}) and
\begin{equation}\label{eq:Wresult}
W^{(0)}(\Delta,P_+) = \frac{G_F^2\,H_\Gamma(\mu_h)}{192 \pi^3}
\frac{m_b^3}{(M_B-P_+)^3}
\int\limits_0^{\Delta-P_+} dk\;F(P_++k,k,\mu_h,\mu_i)\;,
\end{equation}
where
\begin{equation}\label{eq:Fdef}
F(P',k,\mu_h,\mu_i) = (M_B-P')^5 \int\limits_0^{y_{\rm max}[P']}dy
\int\limits_0^{\e_{\rm max}[P',y]} d\e\; y^{-2a_\Gamma(\mu_h,\mu_i)} 
H_u(y,\e,\mu_h)\, Y(k,\ln y,\mu_i)\;.
\end{equation}
The above two formulas enable us to calculate the weight function in
an automated fashion. The procedure is as follows: first, the
integration limits $\Delta$, $y_{\rm max}[P_+]$, and $\e_{\rm
max}[P_+]$ are specified from the kinematics of the cut. For the next
step it is helpful to decompose $Y(k,\ln y,\mu_i) = \sum_i {\cal
D}_i(k,\mu_i)\, Y_i(\ln y,\mu_i)$, where ${\cal D}_i(k,\mu_i)$ are
distributions in $k$ and independent of $y$. (For example, we will see
below that there are only three different distributions in the
perturbative expansion of $Y(k,\ln y,\mu_i)$ to two-loop order.)
Likewise, we decompose $F(P',k,\mu_h,\mu_i) = \sum_i {\cal
D}_i(k,\mu_i)\, F_i(P',\mu_h,\mu_i)$. It is then straight-forward to
calculate the functions $F_i(P',\mu_h,\mu_i)$ by integrating over $\e$
and $y$ in (\ref{eq:Fdef}). Finally the integration over $k$ is
performed in (\ref{eq:Wresult}), where the distributions ${\cal
D}_i(k,\mu_i)$ act on $F_i(P_++k,\mu_h,\mu_i)$.

Equation~(\ref{eq:Wresult}) together with (\ref{eq:Ydef}) is the main
result of this paper. We stress that both formulas are exact
factorization theorems (in the convolution sense), valid to all orders
in perturbation theory, and (\ref{eq:Wresult}) gives the exact
leading-power weight function for arbitrary cuts.

\subsection{Perturbative calculation}

The computation of the kernel $Y(k,\ln y,\mu)$ in (\ref{eq:Ydef})
requires the knowledge of the jet integral $j(L,\mu)$. The jet
function $J(p^2,\mu_i)$ has been calculated to one-loop order
explicitly \cite{Bauer:2003pi,Bosch:2004th}, and the non-constant part
of $j(L,\mu)$ at two-loop order can be extracted from RG evolution
\cite{Neubert:2005nt}. Because the jet integral is of central
importance to the present work, we find it legitimate to re-derive its
dependence on $L$ in detail. In particular, from the RG equations for
the shape function and the leading-power current in soft-collinear
effective theory (SCET, see also \cite{Bauer:2000yr}),
\begin{eqnarray}
0 &=& \int_0^{\hat\omega} d\hat\omega' \left\{ \left[ \frac{d}{d\ln\mu} 
  + 2 \gamma_S(\mu)\right] \delta(\hat\omega-\hat\omega') 
  - 2 \Gamma_c(\mu) \left( \frac{1}{\hat\omega-\hat\omega'} \right)_*^{[\mu]} 
  \right\} \hat S(\hat\omega,\mu) \;,\\
0 &=& \left[\frac{d}{d\ln\mu} + 2\gamma'(\mu) 
  + 2 \Gamma_c(\mu) \ln\frac{m_b}{\mu} \right] \int_0^{P_+} d\hat\omega\; 
  m_bJ(m_b(P_+-\hat\omega),\mu) \; \hat S(\hat\omega,\mu)\;,\nonumber
\end{eqnarray}
we can derive the RG equation governing $j(L,\mu)$ with 
$L = \ln Q^2/\mu^2$ and arbitrary $Q^2$.  Above, $\Gamma_c(\mu)$ is
the cusp anomalous dimension \cite{Korchemsky:1987wg}, which has been
calculated to three-loop order \cite{Moch:2004pa}, the remaining
anomalous dimensions $\gamma_S(\mu)$ and $\gamma'(\mu)$ are known to
two-loop order \cite{Neubert:2004dd,Korchemsky:1992xv,Gardi:2005yi}. 
(For a definition of the star distribution that acts on the shape
function see (\ref{stardistris}) below.) When combining the two RG
equations, one finds
\begin{equation} 
\frac{d}{d\ln \mu} j(L,\mu) = -2 \left[\Gamma_c(\mu) L 
  + \gamma^J(\mu)\right] j(L,\mu) - 2 \Gamma_c \int_0^1 \frac{dz}{z} 
  \big[j(L+\ln(1-z),\mu) - j(L,\mu) \big],
\end{equation}
with $\gamma^J = \gamma'-\gamma_S$.  This integro-differential
equation can be solved perturbatively by choosing a polynomial ansatz
for $j(L,\mu)$.  As a result the integral in the above equation
leads to the appearance of the Riemann zeta-function
$\zeta_n$. Specifically, after expanding the QCD $\beta$-function and
anomalous dimensions as
\begin{equation}
\beta(\mu) = \frac{d\alpha_s(\mu)}{d\ln\mu} = -2 \alpha_s(\mu) 
  \sum\limits_{n=0}^\infty \beta_n \left( \frac{\alpha_s(\mu)}{4\pi} 
  \right)^{n+1}\;,\quad
\Gamma_c(\mu) = \sum\limits_{n=0}^\infty \Gamma_n 
  \left( \frac{\alpha_s(\mu)}{4\pi} \right)^{n+1}\;,
\end{equation}
and similarly $\gamma^J(\mu)$, we find to two-loop accuracy
\begin{eqnarray}
j(L,\mu) &=& 1 
  + \frac{\alpha_s(\mu)}{4\pi} \left[ 
    b_0^{(1)} + \gamma_0^J L + \frac12 \Gamma_0 L^2 \right] \\
&& \mbox{} + \left( \frac{\alpha_s(\mu)}{4\pi} \right)^2 
  \Bigg[ b_0^{(2)} + \left( b_0^{(1)} (\gamma_0^J-\beta_0) + \gamma_1^J 
  - \frac{\pi^2}{6}\Gamma_0\gamma_0^J + \zeta_3 \Gamma_0^2 \right) L 
  \nonumber \\
&& \mbox{} + \frac12 \left( \gamma_0^J (\gamma_0^J-\beta_0) 
  + b_0^{(1)}\Gamma_0+\Gamma_1-\frac{\pi^2}{6} \Gamma_0^2 \right)L^2 
  - \left( \frac16\beta_0-\frac12\gamma_0^J \right) \Gamma_0 L^3 
  +\frac18\Gamma_0^2 L^4 \Bigg] \nonumber .
\end{eqnarray}
Here, $b_0^{(1)} = C_F (7-\pi^2)$ is the one-loop constant, and the
two-loop constant $b_0^{(2)}$ is currently unknown. To determine it a
multi-loop calculation will be necessary. However, this constant does
not enter the two-loop result for the kernel $Y(k,\ln y,\mu_i)$.

Next, we need to find an ansatz for $Y(k,\ln y,\mu)$ which satisfies
(\ref{eq:Ydef}).  At tree-level $j(L,\mu)=1$ and therefore $Y(k,\ln
y,\mu)=\delta(k)$ is also independent of $y$.  It follows that the
constant $b_0^{(2)}$ cancels in the relation~(\ref{eq:Ydef}) to
two-loop accuracy.  Beyond the tree approximation the integrated
$Y(k,\ln y,\mu)$ (defined equivalently to (\ref{eq:j})) must pick up a
logarithmic dependence on $\Omega$, as can be seen by interchanging
the integrations in (\ref{eq:Ydef}); but $Y(k,\ln y,\mu)$ itself must
not depend on $\Omega$. Objects that accomplish that are already known
from the jet function $J(p^2,\mu)$, and are called
``star-distributions'' \cite{DeFazio:1999sv} (see also
\cite{Bauer:2003pi,Bosch:2004th,Lange:2005yw}). Their definitions are
such that, when integrated over an interval $\Omega$, they act on a
function $\phi(k)$ as
\begin{eqnarray}\label{stardistris}
   \int_0^\Omega dk\,\left(\frac{1}{k} \right)_*^{[\mu^2/m_b]} \phi(k)
   &=& \int_0^{\Omega} dk\,\frac{\phi(k) - \phi(0)}{k}\,
    + \phi(0) \,\ln \frac{m_b \Omega}{\mu^2} \,, \\
   \int_0^\Omega dk\,\left(\frac{1}{k}\ln\frac{m_b k}{\mu^2}
     \right)_*^{[\mu^2/m_b]} \phi(k)
   &=& \int_0^{\Omega} dk\,\frac{\phi(k) - \phi(0)}{k}
       \ln\frac{m_b k}{\mu_i^2} \,
    + \frac{\phi(0)}{2} \,\ln^2 \frac{m_b \Omega}{\mu^2} \,.\nonumber
\end{eqnarray}
Therefore, the ansatz for the jet kernel reads (here with $L=\ln y$
for brevity)
\begin{eqnarray} \label{eq:Yansatz}
Y(k,L,\mu) &=& \delta(k)
+ \frac{\alpha_s(\mu)}{4\pi} \left[ c_0^{(1)}(L)\, \delta(k) + 
    c_1^{(1)}(L)\left( \frac1k \right)^{[\mu^2/m_b]}_* \right] \\
&& \mbox{} + \left( \frac{\alpha_s(\mu)}{4\pi} \right)^2 
  \left[ c_0^{(2)}(L)\, \delta(k) + 
    c_1^{(2)}(L)\left( \frac1k \right)^{[\mu^2/m_b]}_* + 
    c_2^{(2)}(L)\left( \frac1k \ln \frac{m_b k}{\mu^2} 
    \right)^{[\mu^2/m_b]}_* 
  \right]. \nonumber
\end{eqnarray}
In (\ref{eq:Wresult}) it acts on all $k$-dependent objects, i.e., on
the prefactor $(M_B-P_+-k)^2$ and on the integration limits $y_{\rm
max}[P_++k]$ and $\e_{\rm max}[P_++k,y]$. To determine the coefficients
$c^{(n)}_i(L)$, however, we consult (\ref{eq:Ydef}), where the star
distributions act on the $k$-dependent logarithms. We find%
\footnote{The expressions for the coefficients are generalizations of
the corresponding findings in \cite{Lange:2005qn}. As mentioned in
footnote~\ref{ftn:f(k)}, the connection is that the integrations over
$\e$ and $y$ were immediately performed, leading to the replacement of
$L^n \to T_n$ and subsequent devision by $T_0$ in the notation of
\cite{Lange:2005qn}. This was possible because for a pure cut on
$P_+$ one has $\e_{\rm max}=y$ and $y_{\rm max} = 1$ independent of
$P_+$, unlike the general case considered here.}
\begin{eqnarray} \label{eq:cCoeffs}
c^{(1)}_1(L) &=& \Gamma_0 L \;, \quad
c^{(1)}_0(L) = \gamma^J_0 L + \frac{\Gamma_0}{2} L^2 \;, \\
c^{(2)}_2(L) &=& - \Gamma_0 L (\beta_0 - \Gamma_0 L) \;, \quad 
c^{(2)}_1(L) = \left[ \Gamma_1 - \beta_0 \gamma^J_0 \right] L + 
\left[ \Gamma_0 \gamma^J_0 - \beta_0 \frac{\Gamma_0}{2} \right] L^2 + 
\frac{\Gamma_0^2}{2} L^3 \nonumber \;, \\
c^{(2)}_0(L) &=& \left[\gamma^J_1 - b^{(1)}_0 \beta_0 \right] L + 
\left[ \frac{\gamma^J_0}{2}(\gamma^J_0 - \beta_0) + \frac{\Gamma_1}{2} 
  - \frac{\pi^2}{12} \Gamma_0^2 \right] L^2 + 
\frac{\Gamma_0}{2} \left[\gamma^J_0 - \frac13 \beta_0 \right] L^3 +
\frac{\Gamma_0^2}{8}\nonumber L^4 \;.
\end{eqnarray}
At this point we interrupt the discussion briefly and consider the
result for the weight function in (\ref{eq:Wresult}) again. As
mentioned earlier, the $b$-quark mass enters the normalization of the
weight function since we also normalized the photon spectrum. In order
to avoid the renormalon ambiguities of the pole scheme it is favorable
to use a low-scale subtracted quark-mass definition, for which we
adopt the shape-function scheme \cite{Bosch:2004th}. The two mass
definitions are connected via
\begin{equation}\label{eq:mbSF}
m_b^{\rm pole} = m_b^{\rm SF}(\mu_*) 
+ \mu_* \frac{C_F \alpha_s(\mu_h)}{\pi}+ \ldots \;,
\end{equation}
and we will refer to $m_b^{\rm SF}(\mu_*)$ at $\mu_* = 1.5$ GeV as
$m_b$ for brevity, for the rest of this paper. Apart from the
normalization to the weight function, $m_b$ also enters through
radiative logarithms in the hard functions and through star
distributions in the jet kernel. Since the hard functions are only
kept to one-loop order, the above redefinition of $m_b$ does not
affect the result. $Y(k,\ln y,\mu_i)$, on the other hand, is given
to two-loop accuracy in (\ref{eq:Yansatz}). It follows that
$c_0^{(1)}$ receives a contribution at this level, 
\begin{equation}
c^{(1)}_0(L) = \left[ \gamma_0 + \frac{C_F \alpha_s(\mu_h)}{\pi} 
\frac{\mu_*}{m_b} \Gamma_0 \right] L + \frac{\Gamma_0}{2} L^2\;.
\end{equation}

In the remainder of this section we collect the other ingredients of
(\ref{eq:Wresult}) for completeness. The hard function in semileptonic
decays is known to 1-loop order in perturbation theory and given by
\begin{equation}\label{eq:Hu}
\begin{aligned}
    H_u(y,\e,\mu_h) 
=&\; 12 (y-\e)(1-y+\e) \bigg\{ 1+\frac{C_F\alpha_s(\mu_h)}{4\pi} \Big[
    -4 \ln^2 \frac{ym_b}{\mu_h} + 10 \ln\frac{ym_b}{\mu_h} - 4\ln y \\
&  \mbox{} \hspace{30mm} - 4L_2(1-y) - \frac{\pi^2}{6}-12 \Big]\bigg\} 
    - 6 (y-\e) \frac{C_F\alpha_s(\mu_h)}{\pi} \ln y\;.
\end{aligned}
\end{equation}
The hard function for the normalized photon spectrum reads 
\begin{eqnarray}
H_\Gamma(\mu_h) &=& 1 + \frac{C_F\alpha_s(\mu_h)}{4\pi}
    \Bigg\{ 4\ln^2 \frac{m_b}{\mu_h} - 10\ln \frac{m_b}{\mu_h} 
    + 7 - \frac{7\pi^2}{6} + 12 \frac{\mu_*}{m_b} 
    \nonumber \\
&& \mbox{} - 2\ln^2 \delta_* - (7+4\delta_*-\delta_*^2)\ln \delta_* 
    + 10 \delta_* + \delta_*^2 - \frac23 \delta_*^3 \nonumber \\
&& \mbox{} + 
   \frac{[C_1(\mu_h)]^2}{[C^{\rm eff}_{7\gamma}(\mu_h)]^2}\, 
     \hat f_{11}(\delta_*)
   + \frac{C_1(\mu_h)}{C^{\rm eff}_{7\gamma}(\mu_h)} \hat f_{17}(\delta_*)
   + \frac{C_1(\mu_h)\,C_{8g}^{\rm eff}(\mu_h)}
     {[C^{\rm eff}_{7\gamma}(\mu_h)]^2} \hat f_{18}(\delta_*) \nonumber \\
&& \mbox{} + \frac{C_{8g}^{\rm eff}(\mu_h)}
     {C^{\rm eff}_{7\gamma}(\mu_h)} \hat f_{78}(\delta_*)
+ \frac{[C_{8g}^{\rm eff}(\mu_h)]^2}
     {[C^{\rm eff}_{7\gamma}(\mu_h)]^2} \hat f_{88}(\delta_*)
  \Bigg\}.
\end{eqnarray}
Here, $\delta_* = 1 - 2 E_*/m_b = 0.9$, and the functions $\hat
f_{ij}(\delta_*)$ capture effects from operator mixing
\cite{Kagan:1998ym} and can be found in this notation in
\cite{Neubert:2004dd}. Note also the term proportional to
$\alpha_s(\mu_h)\, \mu_*/m_b$, which ensures that the three powers of
$m_b$ in (\ref{eq:Wresult}) are defined in the shape-function scheme.
Finally, the RG function $a_\Gamma$ is defined as the integrated cusp
anomalous dimension from $\ln \mu_i$ to $\ln \mu_h$, which yields
\begin{eqnarray}\label{eq:aGamma}
a_\Gamma(\mu_h,\mu_i) &=& \frac{\Gamma_0}{2\beta_0}\,\Bigg\{
\ln\frac{\alpha_s(\mu_i)}{\alpha_s(\mu_h)} 
+ \frac{\alpha_s(\mu_i) - \alpha_s(\mu_h)}{4\pi} 
\left[ \frac{\Gamma_1}{\Gamma_0}-\frac{\beta_1}{\beta_0} \right] \nonumber\\
&&\mbox{}+ \frac{\alpha_s^2(\mu_i) - \alpha_s^2(\mu_h)}{32\pi^2} 
\left[ \frac{\beta_1}{\beta_0} \left( \frac{\beta_1}{\beta_0} - 
\frac{\Gamma_1}{\Gamma_0} \right) - \frac{\beta_2}{\beta_0} +
\frac{\Gamma_2}{\Gamma_0} \right]+ \dots \Bigg\}.
\end{eqnarray}
When combining the various quantities into (\ref{eq:Wresult}) the
result should be re-expanded in $\alpha_s$ to the order in which we
are working. However, it is convenient to treat
$a_\Gamma(\mu_h,\mu_i)$ as a running ``physical'' quantity (similar to
$\alpha_s(\mu)$), which is not expanded. This is the same approach as
put forward in \cite{Lange:2005qn}, and we will use it in the
phenomenological applications in Section~\ref{sec:examples}.

\paragraph{Expansion coefficients of the anomalous dimensions.}

To three-loop order, the coefficients of the $\beta$ function read
\cite{Tarasov:au}
\begin{eqnarray}
\beta_0 &=& \frac{11}{3}\,C_A - \frac23\,n_f \,, \qquad
\beta_1 = \frac{34}{3}\,C_A^2-\frac{10}{3}\,C_A\,n_f-2C_F\,n_f \,,\\
\beta_2 &=& \frac{2857}{54}\,C_A^3
+ \left( C_F^2 - \frac{205}{18}\,C_F C_A
- \frac{1415}{54}\,C_A^2 \right) n_f
+ \left( \frac{11}{9}\,C_F+\frac{79}{54}\,C_A\right)n_f^2\,,\nonumber
\end{eqnarray}
where $n_f=4$ is the number of light flavors, $C_A=3$ and
$C_F=4/3$. The cusp anomalous dimension to three-loop order is given
by \cite{Korchemsky:1987wg,Moch:2004pa}
\begin{eqnarray}
\Gamma_0 &=& 4C_F \,, \qquad
\Gamma_1 = 8C_F \left[ \left( \frac{67}{18} - \frac{\pi^2}{6} \right) C_A 
  - \frac59\,n_f \right] , \\
\Gamma_2 &=& 16C_F \bigg[ \left(\frac{245}{24} - \frac{67\pi^2}{54}
  + \frac{11\pi^4}{180} + \frac{11}{6}\,\zeta_3 \right) C_A^2
  + \left( - \frac{209}{108} + \frac{5\pi^2}{27} - \frac73\,\zeta_3
  \right) C_A\,n_f \nonumber\\
  &&\mbox{} \qquad + \left( - \frac{55}{24} + 2\zeta_3 \right) C_F\,n_f
  - \frac{1}{27}\,n_f^2 \bigg] . \nonumber 
\end{eqnarray}
We also need the anomalous dimension coefficients for the integrated
jet function $j(L,\mu)$, which are \cite{Neubert:2004dd}
\begin{eqnarray}
\gamma_0^J &=& -3 C_F \\
\gamma_1^J &=& C_F \left[ -\left( \frac32-2\pi^2+24\zeta_3 \right)C_F - 
\left( \frac{1769}{54}+\frac{11}{9}\pi^2-40 \zeta_3 \right)C_A + 
\left(\frac{121}{27}+\frac29 \pi^2 \right)n_f \right]. \nonumber 
\end{eqnarray}

\section{Power corrections}
\label{sec:PowCorr}

Factorization theorems exist at each level of power counting for
differential decay rates in inclusive heavy-quark decays. We
differentiate between two types of power corrections, ``kinematical''
and ``hadronic''. The first class arises simply because we have
restricted our discussion to a particular portion in phase-space,
where $P_+ \ll M_B$. These corrections are power suppressed in the
shape-function region, but are of leading power when integrated over a
domain that is comparable to $M_B$ (OPE region). A different way of
thinking about kinematical corrections is to view them as the
equivalent of the factorization theorems (\ref{eq:Xulnu}) and
(\ref{eq:Xsgamma}) with subleading hard and jet functions. However, no
complete scale separation has been achieved for these power
corrections yet, but the products of subleading hard and jet functions
are known in fixed-order perturbation theory and to all powers from
\cite{DeFazio:1999sv,Kagan:1998ym}. Since kinematical corrections
start at $\alpha_s(\bar\mu)$ and are numerically small for all
prominent cuts, this approximate treatment suffices. The scale
$\bar\mu$ is typically near $\sqrt{m_b\Lambda_{\rm QCD}}$, but
independent of $\mu_i$ and $\mu_h$.

The second class of corrections comes from subleading hadronic
structure functions \cite{Bauer:2001mh,Leibovich:2002ys,Bauer:2002yu,%
Neubert:2002yx,Burrell:2003cf,Lee:2004ja,Bosch:2004cb,Beneke:2004in}.
Already at first subleading order there are multiple independent shape
functions entering the calculation of the differential decay
rates. Furthermore, they appear in different linear combinations in
the semileptonic and radiative cases, so that at this stage a weight
function cannot relate semileptonic decay rates to radiative ones
alone. In equation~(\ref{eq:SFfree}) we have therefore added a second
term labeled $\Gamma_{\rm rhc}$. Note that this term is different from
the subleading shape-function contribution to the semileptonic decay
rate (denoted $\Gamma_u^{\rm hadr}$ in \cite{Lange:2005yw})
since contributions from subleading hadronic corrections to the photon
spectrum are convoluted with the leading-power weight function and
must be subtracted. As a result the residual hadronic power
corrections are collected in $\Gamma_{\rm rhc}$. Currently the hard
and jet functions in the subleading shape-function contributions to
the decay rates are only known at tree-level.

\subsection{Kinematical corrections}

Since kinematical corrections come with the leading shape function,
the corresponding contribution to the weight function $W^{\rm
kin}(\Delta, P_+)$ can be calculated without the introduction of any
hadronic uncertainty. These terms start at order $\alpha_s(\bar\mu)$
and we only compute the correction to this order.  It thus follows
that $W^{\rm kin} \otimes d\Gamma_s^{(0)} = \Gamma_u^{\rm kin} -
W^{(0)} \otimes d\Gamma_s^{\rm kin}$, symbolically. (Here and below
the abbreviation $d\Gamma_s$ in statements within the main text
denotes the {\em normalized} photon spectrum.) The relevant
expressions for the differential decay rates have been collected in
\cite{Lange:2005yw} and may be written as (here and below $a_\Gamma =
a_\Gamma(\mu_h,\mu_i)$ for brevity)
\begin{eqnarray}
\Gamma_u^{\rm kin} \Big|_{\rm cut}
&=& N_u \int\limits_0^\Delta dP_+ \int\limits_0^{y_{\rm max}[P_+]}dy 
    \int\limits_0^{\e_{\rm max}[P_+,y]}d\e\;y^{-2a_\Gamma} (M_B-P_+)^4 
    \int\limits_0^{P_+}d\hat\omega\; K_u(x,y,\e)\, \hat S(\hat\omega)\;,
    \nonumber \\
\frac{1}{\Gamma_s}\frac{d\Gamma_s^{\rm kin}}{dP_+}
&=& N_s \frac{(M_B-P_+)^2}{m_b^3} 
    \int\limits_0^{P_+}d\hat\omega\; K_s(x)\, \hat S(\hat\omega)\;, \qquad
    \hbox{with}\;\; x = \frac{P_+-\hat\omega}{M_B-P_+}\;,
\end{eqnarray}
where we have collected all terms in the functions $K_u(\e,y,x)$ and
$K_s(x)$, and abbreviated the different norms by $N_u$ and $N_s$. The
leading weight function at tree-level is taken from
(\ref{eq:Wresult}). We now transform the expression for $\Gamma_u^{\rm
kin} - W^{(0)} \otimes d\Gamma_s^{\rm kin}$ in three steps: First, we
interchange the order of the $P_+$ and $\hat\omega$
integrations. Next, the integration variable $P_+$ is substituted by
$k = P_+-\hat\omega$. Finally the variable $\hat\omega$ is renamed by
$P_+$. After this has been done (no manipulation is needed for the
term $W^{\rm kin} \otimes d\Gamma_s^{(0)}$), we find
\begin{eqnarray}
W^{\rm kin}(\Delta,P_+) 
&=& \frac{N_u}{N_s} \frac{m_b^3}{(M_B-P_+)^3}
    \int\limits_0^{\Delta-P_+}dk\; (M_B-P_+-k)^4 
    \int\limits_0^{y_{\rm max}[P_++k]}dy 
    \int\limits_0^{\e_{\rm max}[P_++k,y]}d\e\;y^{-2a_\Gamma} \nonumber \\
&&\mbox{} \times
    \left[ K_u\left(\frac{k}{M_B-P_+-k},y,\e\right) 
    - H_u^{(0)}(y,\e)\,K_s\left(\frac{k}{M_B-P_+-k}\right) \right],
\end{eqnarray}
where $H_u^{(0)}(y,\e)$ denotes the tree-level part of the hard
function in (\ref{eq:Hu}). 

We are now going to restrict the calculation to the first-order power
corrections because of mixing effects with hadronic power corrections
at higher orders. For example, the second-order power correction to
the weighted integral over the photon spectrum contains a term $W^{\rm
kin(1)} \otimes d\Gamma_s^{\rm hadr(1)}$. (As before, the superscript
denotes the order in power counting.) For such a term we would require
a compensation in the residual correction $\Gamma_{\rm rhc}$ at order
$\alpha_s(\bar\mu)$, which goes beyond the scope of this paper.
Including only first-order power corrections is not a bad
approximation; the studies in \cite{Lange:2005yw} have shown that the
full kinematical corrections can be approximated very well by
including the first term in the power expansion only. At this level
the functions $K_u(x,y,\e)$ and $K_s(x)$ involve only two different
functional dependences on $x$, a constant and a term proportional to
$\ln x$. We find
\begin{equation}
\begin{aligned}\label{eq:WkinResult}
W^{\rm (kin,1)}(\Delta, P_+) 
= & \frac{G_F^2}{192\pi^3} \frac{m_b^3}{(M_B-P_+)^3} 
  \frac{C_F\alpha_s(\bar\mu)}{4\pi} 
  \int\limits_0^{\Delta-P_+} dk\;(M_B-P_+-k)^4 \\
& \mbox{} \times \int\limits_0^{y_{\rm max}[P_++k]} dy
  \int\limits_0^{\e_{\rm max}[P_++k,y]} d\e \;y^{-2a_\Gamma}
  \left[ A(y,\e) + B(y,\e) \ln\frac{k}{M_B-P_+-k} \right]
\end{aligned}
\end{equation}
with
\begin{eqnarray}
  A(y,\e) 
&=& 12 \Bigg[ \frac{\e(5+27\e)}{y} - 4 - 53\e-25\e^2 
    + y(25+46\e) - 21 y^2 \nonumber \\
&&  \mbox{} + 4\,\frac{y-\e}{y} \left[ 3+5\e -(4+\e)y +y^2 \right]\ln y 
    \Bigg] \nonumber \\
&&  \mbox{} + 12 (y-\e)(1-y+\e) \Bigg[ 
    \left(\frac13 - \frac49 \ln \frac{m_b}{m_s} \right) 
    \frac{[C_{8g}^{\rm eff}(\mu_h)]^2}{[C_{7\gamma}^{\rm eff}(\mu_h)]^2} - 
    \frac{10}{3} \frac{C_{8g}^{\rm eff}(\mu_h)}{C_{7\gamma}^{\rm eff}(\mu_h)} 
    \nonumber \\
&&  \mbox{} + \frac83 \left( 
    \frac{C_1(\mu_h)}{C_{7\gamma}^{\rm eff}(\mu_h)}-
    \frac13 \frac{C_1(\mu_h) C_{8g}^{\rm eff}(\mu_h)}{[C_{7\gamma}^{\rm eff}
    (\mu_h)]^2} \right) \,g_1(z) - \frac{16}{9}\,
    \frac{[C_1(\mu_h)]^2}{[C_{7\gamma}^{\rm eff}(\mu_h)]^2}\,g_2(z) 
    \Bigg], \nonumber \\
    B(y,\e) 
&=& 48 \, \frac{y-\e}{y} (1-y)(-3+5y-5\e) - 
    \frac83 (1-y+\e)(y-\e) \frac{[C_{8g}^{\rm eff}(\mu_h)]^2}
    {[C_{7\gamma}^{\rm eff}(\mu_h)]^2} \;. \nonumber 
\end{eqnarray}
In these expressions $C_i(\mu_h)$ denote the (effective) Wilson
coefficients of the relevant operators in the effective weak 
Hamiltonian for $\bar B\to X_s\gamma$ decay. Charm-loop penguin
contributions to the hard function of the photon spectrum are encoded
in the functions $g_1(z)$ and $g_2(z)$, which depend on the variable
$z=(m_c/m_b)^2$ \cite{Kagan:1998ym}. They are
\begin{equation}
   g_1(z) = \int_0^1dx\,x\,\mbox{Re} \left[\,
    \frac{z}{x}\,G\left(\frac{x}{z}\right) + \frac12 \,\right] , \nonumber\\
\end{equation}
\begin{eqnarray}
   g_2(z) &=& \int_0^1dx\,(1-x) \left|\,\frac{z}{x}\,
    G\left(\frac{x}{z}\right) + \frac12\,\right|^2 \,, \\
   G(t) &=& \left\{ \begin{array}{ll}
    -2\arctan^2\sqrt{t/(4-t)} & ~;~ t<4 \,, \\[0.1cm]
    2 \left( \ln\Big[(\sqrt{t}+\sqrt{t-4})/2\Big]
    - \displaystyle\frac{i\pi}{2} \right)^2 & ~;~ t\ge 4 \,.
   \end{array} \right. \nonumber
\end{eqnarray}
This concludes the calculation of the weight function.

\subsection{Residual hadronic corrections}
\label{sec:ResHadr}

There are four different hadronic structures entering the first-order
power corrections to the differential decay rates at tree
level. Following \cite{Bosch:2004cb} we denote them by $(\bar\Lambda -
\hat\omega)\hat S(\hat\omega)$, $\hat t(\hat\omega)$, $\hat
u(\hat\omega)$, and $\hat v(\hat\omega)$. The first one in this list
involves the leading shape function and the heavy-quark parameter
$\bar\Lambda$. This parameter ensures that $(\bar\Lambda -
\hat\omega)\hat S(\hat\omega)$ has a vanishing norm, as all subleading
shape functions must have. It is possible to absorb the effect of this
function into a weight-function contribution, which in turn only
depends on $(\bar\Lambda - P_+)$; however, because of the zero-norm
constraint it is difficult%
\footnote{In \cite{Lange:2005qn} a specific default model for
subleading shape functions was adopted such that this problem is
avoided. A default model of this kind exists -- but is different --
for each kinematical cut.}  to assign the correct numerical value of
$\bar\Lambda$ in that case. Instead, we keep this contribution
together with the subleading shape functions in $\Gamma_{\rm rhc}$. A
straight-forward calculation yields
\begin{eqnarray}\label{eq:resSSF}
    \Gamma_{\rm rhc} \Big|_{\rm cut}
&=& \frac{G_F^2\,U(\mu_h,\mu_i)}{16\pi^3} \int_0^\Delta dP_+\;
    (M_B-P_+)^4 \int\limits_0^{y_{\rm max}[P_+]}dy 
    \int\limits_0^{\e_{\rm max}[P_+,y]}d\e\;y^{-2a_\Gamma}\; 
    \frac{y-\e}{y}\nonumber \\
&& \mbox{} \times \Bigg\{ 2(1-y)(y-\e)\;(\bar\Lambda - P_+) \hat S(P_+)
   + 2(1-y) \left( y-\e + \frac{\e}{y} \right) \hat t(P_+) \\
&& \mbox{} \quad + (1-y)(1-y+\e)\, \hat u(P_+) + 
   \left( 1+y-\frac{y-\e}{y} (2-y+y^2)\right) \hat v(P_+) \nonumber \\
&& \mbox{} \quad + y(1-y+\e)\,m_s^2 \hat S'(P_+) \Bigg\}, \nonumber 
\end{eqnarray}
where we have kept the RG-evolution function $y^{-2a_\Gamma}$, as well as
$U(\mu_h,\mu_i)$ \cite{Lange:2005yw}. Its structure is such that
\begin{equation}\label{eq:U}
\ln U(\mu_h,\mu_i) = 2 S_\Gamma(\mu_h,\mu_i) - 2 a_\Gamma(\mu_h,\mu_i) 
\ln \frac{m_b}{\mu_h} - 2 a_{\gamma'}(\mu_h,\mu_i)\;,
\end{equation}
and we evaluate to leading-order the Sudakov exponent 
(with $r = \alpha_s(\mu_i)/\alpha_s(\mu_h)$)
\begin{equation}
S_\Gamma(\mu_h,\mu_i)
= \frac{\Gamma_0}{4\beta_0^2}\,\left\{
\frac{4\pi}{\alpha_s(\mu_h)} \left( 1-\frac1r-\ln r \right)
+ \left( \frac{\Gamma_1}{\Gamma_0} - \frac{\beta_1}{\beta_0} \right)
(1-r+\ln r) + \frac{\beta_1}{2\beta_0} \ln^2 r \right\},
\end{equation}
and the RG function $a_{\gamma'} = \gamma'_0/(2\beta_0)\,\ln r$
analogous to (\ref{eq:aGamma}). The use of the anomalous dimension
$\gamma'_0=-5 C_F$ of the leading SCET current is a good approximation
because most of the subleading operators in SCET are build from the
leading SCET current and thus share the same anomalous dimension. 
(A complete resummation requires knowledge of the anomalous dimension
matrix of all subleading operators in SCET. For some discussion, see
e.g.~\cite{Hill:2004if,Arnesen:2005nk,Trott:2005vw}.)

We have included a small correction from a finite $s$-quark mass,
leading to the expression proportional to $m_s^2 S'(P_+)$. The
appearance of hadronic structure functions in (\ref{eq:resSSF})
introduces some irreducible uncertainties in phenomenological
applications of our results. While in principle the leading shape
function $\hat S(\hat\omega)$ can be extracted from the photon
spectrum, see e.g.~\cite{Lange:2005yw}, the forms of the shape
functions $\hat t(\hat\omega)$, $\hat u(\hat\omega)$, and $\hat
v(\hat\omega)$ are unknown. What we know is that their norms vanish
and their first moments are given in terms of the heavy-quark
parameters $\mu_\pi^2$ and $\lambda_2$. In order to estimate effects
from higher moments we define functions $h_n(\hat\omega)$,
$n\in\{t,u,v\}$, via
\begin{equation}\label{eq:SSFmodel}
\hat t(\hat\omega) = -\lambda_2\;\hat S'(\hat\omega)+h_t(\hat\omega)\;,\quad
\hat u(\hat\omega) = -2\; \frac{\mu_\pi^2}{3}\;\hat S'(\hat\omega)
  +h_u(\hat\omega) \;,\quad
\hat v(\hat\omega) = \lambda_2\;\hat S'(\hat\omega)+h_v(\hat\omega)\;.
\end{equation}
As long as each of the $h_n(\hat\omega)$ have zero norm and first
moment the constraints on the subleading shape functions are
satisfied.

Let us now adopt a default model for the terms in
(\ref{eq:resSSF}). For simplicity we use for $\hat S(\hat\omega)$ the
familiar exponential-type functional form
\begin{equation}
\hat S(\hat\omega) = \frac{1}{\bar\Lambda}\frac{b^b}{\Gamma(b)}
\left( \frac{\hat\omega}{\bar\Lambda} \right)^{b-1} 
\exp \left( - b \,\frac{\hat\omega}{\bar\Lambda} \right) ,
\qquad \hbox{with}\quad b = \frac{3\bar\Lambda^2}{\mu_\pi^2}\;,
\end{equation}
with parameters $\bar\Lambda$ and $\mu_\pi^2$. By construction the
moment constraints on $(\bar\Lambda-\hat\omega) \hat S(\hat\omega)$
are respected. The default model for (\ref{eq:SSFmodel}) is defined as
$h_n(\hat\omega)=0$ in all three cases.

In order to estimate the uncertainty introduced by adopting a specific
model we closely follow the procedure of
\cite{Lange:2005yw,Lange:2005qn}, where four different functions
$h_i(\hat\omega)$, $i=1,\ldots, 4$, were constructed with vanishing
norm and first moment. A variation of the functional form of the
subleading shape functions was then achieved by setting
$h_n(\hat\omega) \to \pm h_i(\hat\omega)$, where $n \in \{t,u,v\}$ and
$i \in \{1,\ldots, 4\}$. Combinatorially this means we have $9^3 =
729$ different models for the set of subleading shape functions. The
estimator for the hadronic uncertainty is the maximal deviation from
the default result when sampling over all models.

\section{Examples and Discussion}
\label{sec:examples}

Let us demonstrate the phenomenological implications of the result
(\ref{eq:Wresult}), (\ref{eq:WkinResult}), and (\ref{eq:resSSF}) by
applying the main relation (\ref{eq:SFfree}) for a few examples of
typical kinematic cuts. To disentangle theoretical uncertainties from
experimental ones we are going to pretend that both the normalized
photon spectrum and the semileptonic partial decay rate were measured
with no uncertainty. In particular we consider the photon spectrum
given as
\begin{equation}
  \frac{1}{\Gamma_s(E_*)}\frac{d\Gamma_s}{dP_+} 
= \frac{1}{\Lambda_\gamma}\frac{b_\gamma^{b_\gamma}}{\Gamma(b_\gamma)} 
\left( \frac{P_+}{\Lambda_\gamma} \right)^{b_\gamma-1} 
\exp \left( - b_\gamma \,\frac{P_+}{\Lambda_\gamma} \right) ,
\end{equation}
with $\Lambda_\gamma = 0.77$ GeV and $b_\gamma=2.5$. This model
describes the experimental data by BaBar \cite{Aubert:2005cb}
and Belle \cite{Abe:2005cv} reasonably well.  Furthermore we
need as inputs the heavy-quark parameters $\lambda_2 = 0.12$ GeV$^2$,
$\mu_\pi^2 = (0.25 \pm 0.10)$ GeV$^2$, and the quark masses $m_b =
(4.61\pm 0.06)$ GeV \cite{Neubert:2005nt,Lange:2005yw,Neubert:2004sp},
$m_s = (90\pm 25)$ MeV \cite{Aubin:2004ck,Gamiz:2004ar}, and $m_c/m_b
= 0.222\pm0.027$ \cite{Neubert:2004dd}. Here $m_b$ and $\mu_\pi^2$ are
defined in the shape-function scheme at a scale $\mu_* = 1.5$ GeV,
while $m_s$ is evaluated in the $\overline{\rm MS}$ scheme at $1.5$
GeV. The ratio $m_c/m_b$ is also evaluated in the $\overline{\rm MS}$
scheme, where it is scale invariant. For the strong coupling
$\alpha_s(\mu)$ we use three-loop running from $\alpha_s(M_Z) =
0.1187$ down to $4.25$ GeV, apply matching corrections onto a 4-flavor
theory, and then run to $\mu_h$, $\mu_i$, or $\bar\mu$. Default values
for these scales are taken to be $m_b/\sqrt2$, $1.5$ GeV, and $1.5$
GeV, respectively. To assign a perturbative error, the scales are
varied around these default settings by factors of $\sqrt2$ and
$1/\sqrt2$. In all cases considered below, the analysis of
uncertainties closely follows \cite{Lange:2005qn}, with
similar outcome.

\begin{figure}
\epsfig{file=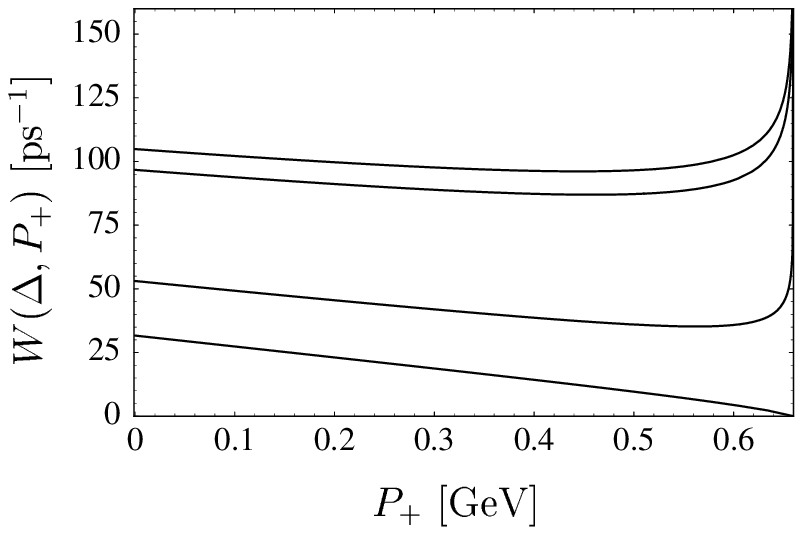, width=8cm}\hspace{5mm}
\epsfig{file=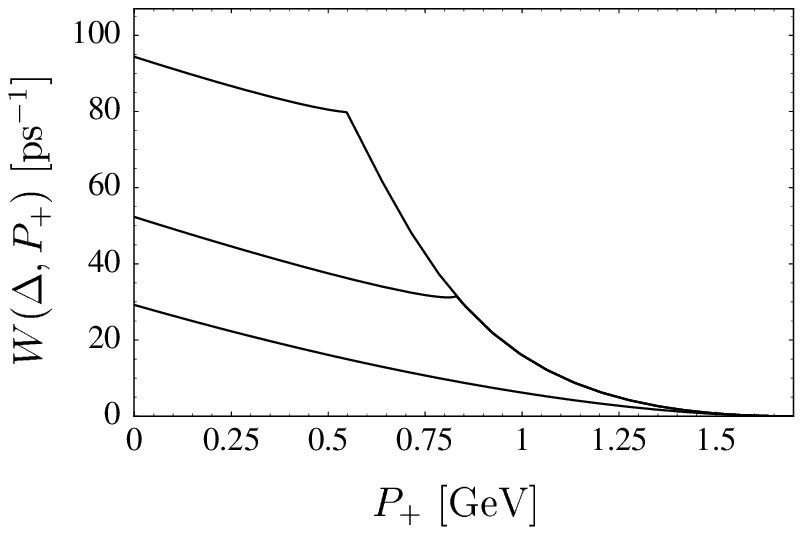, width=8cm}
\caption{\label{fig:examples}Examples of the weight function for
different kinematic cuts. LEFT: Cutting on $P_+\le \Delta = 0.66$ GeV
and $E_l > E_0$. From top to bottom the four functions are for $E_0 =
0$, $E_0 = 1$ GeV, $E_0 = 2$ GeV, and $E_0 = (M_B-\Delta)/2$. RIGHT:
Cutting on $M_X\le M_0 = 1.7$ GeV, $q^2 > q_0^2$, and $E_l > 1$
GeV. The three functions are for $q_0^2 = 0$ (top), $q_0^2 = 8$
GeV$^2$ (middle), and $q_0^2 = (M_B-M_0)^2$ (bottom).}
\end{figure}

\paragraph{\boldmath Cutting on $P_+$ and $E_l$.}

First, let us consider a cut on $P_+ \le \Delta$ with $\Delta = 0.66$
GeV together with a cut on $E_l > E_0$. In terms of the kinematic
variables $y$ and $\e$ this means that
\begin{equation}
y_{\rm max}[P_+] = 1\;, \qquad 
\e_{\rm max}[P_+,y] = {\rm min}\left(1-\frac{2E_0}{M_B-P_+}, y \right).
\end{equation}
Using the central values of the input parameters and scales, the
resulting weight functions are depicted on the left-hand side of
Figure~\ref{fig:examples} for a few examples of $E_0$. There is an
integrable singularity at the endpoint $P_+ \to \Delta$
\cite{Hoang:2005pj,Lange:2005qn} if the lepton cut $E_0$ is small. In
the limit $E_0 \to (M_B - \Delta)/2$, corresponding to a pure cut on
lepton energy, the weight function vanishes at the endpoint and the
singularity disappears.

We now study the case $E_0 = 1$ GeV in more detail. This is a useful
example because the partial branching fraction for this particular cut
has already been measured \cite{Bizjak:2005hn}. First, we investigate
the perturbative uncertainty of the right-hand side of
(\ref{eq:SFfree}), which is obtained by studying the residual scale
dependence entering via the weight function. (It is clear that the
relevant quantity is the entire integral, and not the values of the
weight function for individual values of $P_+$.)  We observe that the
sum of the weighted integral over the photon spectrum $W\otimes
d\Gamma_s = 44.58$ ps$^{-1}$ and the residual hadronic corrections
$\Gamma_{\rm rhc} = -5.65$ ps$^{-1}$ is very stable under scale
variations, but not each of the two terms alone. The NNLO
approximation for the weight function introduces roughly a $4\%$
uncertainty. The error from the LO approximation to the power
suppressed corrections is numerically of equal magnitude, but tends to
cancel a large portion of the scale sensitivity. This leads us to
interpret the perturbative error on the convolution integral as also
the perturbative uncertainty on the sum of both terms, to avoid
counting this error twice. The hadronic uncertainty on the residual
term $\Gamma_{\rm rhc}$ is obtained by taking the maximal deviation
from the central value when sampling over a large set of models for
the subleading shape functions, as outlined in
Section~\ref{sec:ResHadr}. Finally we also vary the numerical value of
$m_b$ and the remaining input parameters $m_s$, $m_c$, and $\mu_\pi^2$
within their stated errors. This yields
\begin{equation}
\frac{W \otimes d\Gamma_s + \Gamma_{\rm rhc}}{\rm ps^{-1}} 
\Bigg|_{\hbox{\scriptsize $\begin{array}{l}
P_+ \le 0.66\;{\rm GeV}, \\
E_l \ge 1\;{\rm GeV}\end{array}$}}
= 38.93 \; ^{+2.23}_{-1.96}\,\hbox{\scriptsize [pert.]}
        \; \pm 1.42 \,\hbox{\scriptsize [hadr.]}
        \; ^{+1.71}_{-1.67} \,\hbox{\scriptsize [$m_b$]}
        \; ^{+0.46}_{-0.63} \,\hbox{\scriptsize [pars.]}\;.
\end{equation}
Further uncertainty enters in practice, because the photon spectrum
cannot be measured over the entire range, but only over a certain
window around the endpoint $E_\gamma = M_B/2$. The normalized photon
spectrum is then obtained using theoretical information on the
fraction of events that fall into this window. The current precision
for these fractions is about $6\%$ \cite{Neubert:2004dd}, which
impacts $W \otimes d\Gamma_s$ directly.  If we further assume that the
left-hand side of relation~(\ref{eq:SFfree}) was given with no
experimental uncertainty, we can extract $|V_{ub}|$ with only a
theoretical error. For example, let us take the central value $Br(P_+
\le 0.66\;{\rm GeV}, E_l \ge 1\;{\rm GeV}) = 1.1\cdot 10^{-3}$
\cite{Bizjak:2005hn}, and dismiss the experimental error. Using the
average lifetime $\tau_B = 1.60\;{\rm ps}$ of the $B$ meson, and
taking the normalization uncertainty on the photon spectrum into
account, we find
\begin{equation}
|V_{ub}| = \left( 4.20\;^{+0.24}_{-0.21} \,
\hbox{\scriptsize [theory]} \right) \cdot 10^{-3}\;.
\end{equation}

\paragraph{\boldmath Cutting on $M_X$, $q^2$, and $E_l$.}

When cutting on $M_X\le M_0$, $q^2 \ge q_0^2$, and $E_l\ge E_0$, the
maximal value of $P_+$ is given by 
$\Delta = {\rm min}(M_0,M_B-\sqrt{q_0^2},M_B-2E_0)$. The phase space
is such that
\begin{eqnarray}
y_{\rm max}[P_+] &=& {\rm min}\left(
  1-\frac{q_0^2}{(M_B-P_+)^2}, 
  \frac{M_0^2 - P_+^2}{P_+(M_B-P_+)} \right), \\
\e_{\rm max}[P_+,y] &=& {\rm min}\left(
1-\frac{2E_0}{M_B-P_+}, y \right). \nonumber
\end{eqnarray}
The right-hand side of Figure~\ref{fig:examples} shows three different
weight functions. In all cases $M_0 = 1.7$ GeV, close to the optimal
value $M_D$, and $E_0 = 1$ GeV as before. From top to bottom the
three curves are for $q^2 = 0$ (pure $M_X$ cut), $q^2 = 8$ GeV$^2$
(mixed cut), and $q^2 = (M_B-M_0)^2$ (pure $q^2$ cut). Note that the
integrable singularity of the pure $P_+$ cut in the previous
discussion is gone, because this point is no longer the
endpoint. Instead, the weight function has a kink, which is expected
from considerations of the phase space depicted in
Figure~\ref{fig:pspace}. The endpoint $\Delta$ is much larger in this
case, and the weight function vanishes there, so that we do not
encounter any singularity anymore. 

As a second example of a $|V_{ub}|$ determination we consider the case
$M_0 = 1.7$ GeV, $q_0^2 = 0$, $E_0 = 1$ GeV, i.e., the top curve in
the plot on the right-hand side of Figure~\ref{fig:examples}. Because
of phase-space restrictions the weight function has a kink at 
$P_+ \approx 0.55$ GeV. For the central values and default models we
find $W\otimes d\Gamma_s + \Gamma_{\rm rhc} = (49.77 - 1.63)$
ps$^{-1}$, and the analysis of uncertainties yields
\begin{equation}
\frac{W \otimes d\Gamma_s + \Gamma_{\rm rhc}}{\rm ps^{-1}} 
\Bigg|_{\hbox{\scriptsize $\begin{array}{l}
M_X \le 1.7\;{\rm GeV}, \\
E_l \ge 1\;{\rm GeV}\end{array}$}}
= 48.14 \; ^{+1.60}_{-1.82}\,\hbox{\scriptsize [pert.]}
        \; \pm 0.47 \,\hbox{\scriptsize [hadr.]}
        \; ^{+1.93}_{-1.88} \,\hbox{\scriptsize [$m_b$]}
        \; ^{+1.18}_{-0.99} \,\hbox{\scriptsize [pars.]}\;.
\end{equation}
Again, we must also add a $6\%$ uncertainty to the norm of the photon
spectrum.  From the input 
$Br(M_X \le 1.7\;{\rm GeV}, E_l \ge 1\;{\rm GeV}) = 1.24\cdot 10^{-3}$
\cite{Bizjak:2005hn} follows
\begin{equation}
|V_{ub}| = \left( 4.01\; ^{+0.18}_{-0.16} \, 
\hbox{\scriptsize [theory]} \right) \cdot 10^{-3} \;.
\end{equation}

The third and last example is the combined cut $M_0 = 1.7$ GeV, 
$q_0^2 = 8$ GeV$^2$, $E_0 = 1$ GeV, whose weight function is shown as
the curve in the middle of the right plot in
Figure~\ref{fig:examples}.  In analogy to the previous cases we obtain
$W\otimes d\Gamma_s + \Gamma_{\rm rhc} = (28.18 - 4.97)$ ps$^{-1}$ and
\begin{equation}
\frac{W \otimes d\Gamma_s + \Gamma_{\rm rhc}}{\rm ps^{-1}} 
\Bigg|_{\hbox{\scriptsize $\begin{array}{l}
M_X \le 1.7\;{\rm GeV}, \\
q^2 \ge 8\;{\rm GeV}^2, \\
E_l \ge 1\;{\rm GeV}\end{array}$}}
= 23.21 \; ^{+1.33}_{-1.51}\,\hbox{\scriptsize [pert.]}
        \; \pm 0.43 \,\hbox{\scriptsize [hadr.]}
        \; ^{+1.10}_{-1.07} \,\hbox{\scriptsize [$m_b$]}
        \; ^{+0.71}_{-0.70} \,\hbox{\scriptsize [pars.]}\;.
\end{equation}
Taking $Br(M_X \le 1.7\;{\rm GeV}, q^2 \ge 8\;{\rm GeV}^2, E_l \ge
1\;{\rm GeV}) = 8.56\cdot 10^{-4}$, which is the mean of the central
values of the measurements reported in
\cite{Bizjak:2005hn,Aubert:2005hb}, leads to a larger value, namely
\begin{equation}
|V_{ub}| = \left( 4.79\; ^{+0.30}_{-0.24} \, 
\hbox{\scriptsize [theory]} \right) \cdot 10^{-3} \;.
\end{equation}

It is of course understood that these values change once an analysis
with the full experimental data is conducted. We believe, however,
that a theoretical error of about $\pm 5\%$ is realistic and consistent
with the previous work in \cite{Lange:2005yw,Lange:2005qn}.

\paragraph{Concluding remarks.} 

In the last section we have given results for the weight function
$W(\Delta,P_+)$ for several commonly employed kinematic cuts. In all
cases the charged-lepton energy was bound to exceed 1 GeV,
which is typically used to identify semileptonic $B$ decays in
practice. We stress that for an additional cut on $P_+$ the photon
spectrum is required over only a small window, where already precise
data exists. For a cut on the hadronic invariant mass, on the other
hand, the photon spectrum is also needed in a regime where its
measurement is very difficult. Cutting away further events in the low
$P_+$ region by virtue of an additional restriction on the leptonic
invariant mass $q^2$ worsens the situation even more, because the
relative importance of the high $P_+$ region is enhanced. It is
therefore expected that the first example, the cut on $P_+$, will
ultimately lead to the most precise determination of $|V_{ub}|$.

In summary, we have presented a formula in (\ref{eq:Wresult}) which is
based on exact factorization theorems for the differential decay
rates, and that allows for the calculation of weight functions for
arbitrary kinematic cuts. Quantities entering this formula were
evaluated with one-loop precision at the hard scale, and complete
two-loop precision at the intermediate scale, including three-loop
running effects in renormalization-group improved perturbation
theory. To achieve further precision we have also included first-order
power corrections resulting from subleading kinematical and hadronic
contributions. The details of the cut are encoded in three kinematic
quantities, $\e_{\rm max}[P_+,y]$, $y_{\rm max}[P_+]$, and
$\Delta$. Once they are specified, one only needs to carry out a few
integrations that lead directly to the weight function.

The use of relations such as (\ref{eq:SFfree}) circumvents the
necessity for fitting models of the leading shape function to the
$\bar B\to X_s\gamma$ photon spectrum and allows for the determination
of $|V_{ub}|$ in a model-independent way at leading power.  The quest
for a precision measurement of $|V_{ub}|$ requires a variety of
different approaches. The results of this paper represent an
alternative route to direct theoretical predictions of partial decay
rates.

\vspace{1.3cm}
\noindent
{\em Acknowledgments:} I would like to thank Matthias Neubert for
comments on the manuscript. I also thank Gil Paz for his collaboration
in the early stages of this work and his comments on the
manuscript. This work was supported in part by funds provided by the
U.S.~Department of Energy (D.O.E.) under cooperative research
agreement DE-FC02-94ER40818.

\end{document}